\documentclass[apjl]{emulateapj}
\usepackage{amsfonts,amsmath,graphicx,natbib,apjfonts,subfigure}

\def\unife{1}
\def\iasfbo{2}
\def\cfa{3}
\shorttitle{Waiting times in GRBs}
\shortauthors{Guidorzi et al.}

\begin{document}
\title{A common stochastic process rules gamma--ray burst prompt emission and X--ray flares}

\author{
C.~Guidorzi\altaffilmark{\unife},
S.~Dichiara\altaffilmark{\unife},
F.~Frontera\altaffilmark{\unife,\iasfbo},
R.~Margutti\altaffilmark{\cfa},
A.~Baldeschi\altaffilmark{\unife},
L.~Amati\altaffilmark{\iasfbo}
}

\altaffiltext{\unife}{Department of Physics and Earth Sciences, University of Ferrara,
  via Saragat 1, I--44122, Ferrara, Italy}
\altaffiltext{\iasfbo}{INAF--IASF Bologna, via Gobetti 101, I--40129, Bologna, Italy}
\altaffiltext{\cfa}{Harvard-Smithsonian Center for Astrophysics, 60 Garden St., Cambridge,
  MA 02138, USA.}

\begin{abstract}
Prompt $\gamma$--ray and early X--ray afterglow emission in gamma--ray bursts (GRBs)
are characterized by a bursty behavior and are often interspersed with long quiescent times.
There is compelling evidence that X--ray flares
are linked to prompt $\gamma$--rays. However, the physical mechanism that leads to the
complex temporal distribution of $\gamma$--ray pulses and X--ray flares is not understood.
Here we show that the waiting time distribution (WTD) of pulses and flares exhibits a
power--law tail extending over 4 decades with index $\sim2$
and can be the manifestation of a common time--dependent Poisson process.
This result is robust and is obtained on different catalogs.
Surprisingly, GRBs with many ($\ge8$) $\gamma$--ray pulses are very unlikely to
be accompanied by X--ray flares after the end of the prompt emission ($3.1\sigma$
Gaussian confidence).
These results are consistent with a simple interpretation: an hyperaccreting disk breaks
up into one or a few groups of fragments, each of which is independently accreted
with the same probability per unit time. Prompt $\gamma$--rays and late
X--ray flares are nothing but different fragments being accreted at the beginning and
at the end, respectively, following the very same stochastic process and likely
the same mechanism.
\end{abstract}


\section{Introduction}
\label{sec:Intro}
The first electromagnetic messenger of a gamma--ray burst (GRB)
is the so--called $\gamma$--ray prompt emission, followed by the early X--ray
afterglow on a timescale from minutes to hours.
Long duration ($>2$--$3$~s) GRBs are nowadays known to be associated with
the core collapse of some kind of massive stars rid of hydrogen envelopes (see
\citealt{Woosley06,HjorthBloom12} for reviews).
Prompt $\gamma$--rays (with energies in the keV--MeV range) are observed
within a given GRB as a sequence of pulses (typically a few up to several dozens).
In addition, for a sizable fraction of GRBs the following decaying
X--ray emission, which marks the end of the $\gamma$--rays,
is characterized by the presence of X--ray flares which are sometimes
observed as late as $10^5$~s \citep{Burrows05b,Chincarini07,Falcone07,Curran08,Bernardini11}.
Although mounting evidence exists that X--ray flares, like $\gamma$--ray pulses,
result from the GRB inner engine activity rather than from external shocks
\citep{LazzatiPerna07,Chincarini10,Margutti10b}, key questions remain unanswered:
what radiation process(es)? What information on the inner engine can we extract?
Is there a common process ruling inner engine activity across several decades in time?

As a matter of fact, both emissions represent a temporal
point process, i.e. a time series characterized by the discrete occurrence of
impulsive events superposed on a continuum. Intense bursting periods
are often interspersed with relatively long (several up to tens of seconds) intervals
with very low activity, compatible with the detector background, which are often
referred to as quiescent times (QTs; \citealt{RamirezRuiz01a,NakarPiran02a,Quilligan02,DragoPagliara07}).
The study of the waiting time distribution (WTD), i.e. of how time intervals between
adjacent peaks distribute, provides clues on the nature of the stochastic process.

In particular, it reveals the degree of memory and correlation and constrains
the physical process responsible for the discontinuous and bursty release of energy.

Processes showing similar on--off intermittency, or, equivalently, bursty behavior
or clusterization, can be found in many fields \citep{Platt93}.
The corresponding WTDs often show power--law tails at long waiting times (WTs),
whose index depends on the degree of clusterization of the time series.
Examples encompass the aftershock sequence observed in earthquakes, described by
Omori's law \citep{Utsu61}, neuronal firing activity, as well as a wide range of
dynamical systems of human activity, such as mail and email exchanges \citep{Oliveira05,Eckmann04},
phone calls (\citealt{Karsai12} and references therein) all the way to violent conflicts \citep{Picoli14}.
These processes are often modeled and interpreted in the context of
self--organized criticality (SOC), where a nonlinear dynamical system reaches a
stable critical point in which continuous energy input is released intermittently
through avalanches and in a scale--free way.
SOC naturally predicts power--laws in energy and WT distributions.
See \citet{Aschwanden14} for a recent review on the many areas displaying SOC behavior.

In astrophysics WTDs are studied for many different kinds of sources, such as
outbursting magnetars \citep{Gogus99,Gogus00,Gavril04}, flare stars \citep{Arzner04},
and particularly the activity of the Sun throughout its cycle.
WTDs of solar X--ray flares exhibit power--law tails with indices
in the range $2.0$--$2.4$ across several decades \citep{Boffetta99,Wheatland00},
depending on the class of flares and flux thresholds.
Related bursty emission from the Sun such as coronal mass ejections (CMEs) are found
to show very similar WTDs, whose index ranges from $\sim1.9$ to $\sim3.0$ in low to
high--activity periods of the solar cycle \citep{Wheatland03}. Likewise, WTDs of
solar radio storms \citep{Eastwood10}, of solar energetic particle and of solar electron
events show very similar power--law indices \citep{Li14}.
Such power--law tailed WTDs are usually interpreted as the consequence of a time--varying
Poisson process produced by SOC systems, in which the energy input rate is intermittent
and directly affects the degree of clusterization of flares \citep{Aschwanden10,Li14}.
In this model, the bursty energy release is the result of avalanches produced in active
regions where magnetic flux is twisted by the moving footpoints, leading to a series of
independent magnetic reconnection events and consequent plasma acceleration.
Alternatively to SOC, interpretations in the context of fully developed MHD turbulence have also
been proposed to explain the bursty dynamics and the power--law WTD:
the intermittent character is the result of a nonlinear dynamics
which makes the convective motion of the fluid and magnetic field swing between
laminar and turbulent regimes repeatedly and chaotically \citep{Boffetta99,Lepreti04}.

The WTD between adjacent peaks in GRB $\gamma$--ray prompt emission profiles was found
to be described by a lognormal --which implies some degree of memory-- \citep{Li96} with an
excess at long values due to QTs \citep{NakarPiran02a,Quilligan02,DragoPagliara07}.
However, when the peak detection efficiency is carefully taken into account, it is found
that the intrinsic WTD at short values is also compatible with an exponential, that is
what is expected for a {\em constant} Poisson (thus memoryless) process \citep{Baldeschi15}.
On the other side of the distribution long QTs could either mark the inner engine temporarily
switching off, or result from modulation of the relativistic wind of shells
\citep{RamirezRuiz01b}, or they could be due to a different physical mechanism from that
of short WTs (e.g., \citealt{Drago08}).

In spite of the impressive data that are routinely being acquired in the {\em Swift} era,
little progress has been reported on WTDs in GRBs. Recently, energy and WT
distributions for GRB X--ray flares have been shown to have power--law tails very similar
to what is observed for solar X--ray flares. In particular, the WTD has a power--law
index of $1.8\pm0.2$ \citep{WangDai13}.
These results were interpreted as evidence for SOC possibly driven by magnetic reconnection
episodes triggered in magnetized shells emitted by differentially rotating millisecond
pulsars or, alternatively, by an hyperaccreting disk around a black hole \citep{Popham99}.

Yet, there are several crucial issues which can be tackled with WTDs:
is there additional evidence for a link between prompt $\gamma$--rays and late
X--ray flares?
To what extent do QTs differ from short WTs?
Is it possible to provide a common description of short WTs, QTs, X--ray flares?
What about rest--frame properties?
Is there evidence for memory in GRB engines?
What can be inferred on GRB engines through the WTD study?

In this paper we address these issues through the analysis of
the WTD of GRB prompt peaks for three independent data sets: {\em Swift}/BAT,
{\em CGRO}/BATSE, and {\em Fermi}/GBM. For the {\em Swift} GRBs
which have also been promptly observed with XRT, we present a joint
analysis of $\gamma$--ray peaks and X--ray flares merged together.
Section~\ref{sec:data} describes the data sample selection and
how we modeled the WTDs.
Here we deliberately did not consider the energy distribution of peaks
and flares, because even though our peak search algorithm identified
moderately overlapped pulses, estimating their energy would require
specific assumptions on their temporal structure.
We therefore decided to postpone it for future investigation.
The results, their implications and interpretation are reported
in Sections~\ref{sec:res} and \ref{sec:disc}, respectively.
Hereafter, uncertainties on best fit parameters are given at 90\%
confidence, unless stated otherwise.

\section{Data analysis}
\label{sec:data}
We searched all long--duration $\gamma$--ray light curves with
{\sc mepsa}\footnote{http://www.fe.infn.it/u/guidorzi/new\_guidorzi\_files/code.html}
\citep{Guidorzi15,MEPSA}, a peak search algorithm designed and calibrated to this goal.
The advantage of {\sc mepsa} compared with analogous algorithms such
as the one by \citet{Li96} (LF) is twofold:
\begin{itemize}
\item it has a lower false positive rate.
This is particularly true for the time intervals in which
the signal drops to background between two adjacent activity periods:
in the best cases the LF false positive rate is 3--5$\times10^{-3}$~bin$^{-1}$,
while the {\sc mepsa} one is 1--2$\times10^{-5}$~bin$^{-1}$ \citep{Guidorzi15};
\item it has a higher true positive rate, especially at low signal
to noise ratios ($\sim$4--5).
\end{itemize}
%

\subsection{Sample selection}
\label{sec:sample} 

\subsubsection{{\em Swift}/BAT data}
\label{sec:BATsample} 
We started with the GRBs detected by {\em Swift}/BAT in burst mode from January 2005
to September 2014, collecting 825 GRBs. We extracted the mask--weighted light
curves in the 15--150~keV energy band with a uniform bin time of 64~ms following
the standard procedure recommended by the
BAT team\footnote{http://swift.gsfc.nasa.gov/analysis/threads/bat\_threads.html}
and applied {\sc mepsa}.
We then imposed a minimum threshold of $5\,\sigma$ significance,
which ensures a very low false positive rate ($<10^{-5}$~bin$^{-1}$;
\citealt{Guidorzi15}) and selected the GRBs with at least two peaks.
We then removed from our sample the short duration GRBs (both with and without extended
emission) by crosschecking with the classification provided in the BAT catalog
\citep{Sakamoto11}, as they will the subject of future investigation.
Since this catalog does not include GRBs from 2010, for
these GRBs we used the $T_{90}$ values as published in the BAT refined
GCN circulars regularly published by the BAT team and set a conservative
minimum threshold of $T_{90}>3$~s.
A couple of GRBs detected by BAT exhibited a very long duration which could not
be covered entirely in burst mode: in one case we used the {\em WIND}/Konus light
curve for GRB\,091024 \citep{Virgili13}, while in the case of GRB\,130925A
we used the peak times as they have been obtained by \citet{Evans14} from the
corresponding Konus light curve.
Finally, we ended up with a sample of 418 long GRBs with at least two significant
($>5\,\sigma$) peaks each, totaling 2000 peaks and 1582 WTs.
Hereafter, we refer to this sample as the BAT set.

\subsubsection{{\em CGRO}/BATSE data}
\label{sec:BATSEsample} 
We took the concatenated 64--ms burst data distributed by the BATSE team
\footnote{ftp://cossc.gsfc.nasa.gov/compton/data/batse/ascii\_data/64ms/}.
For each curve we interpolated the background by fitting with polynomials
of up to fourth degree as suggested by the BATSE team (e.g., \citealt{Guidorzi05c}).
Like in the BAT case, we applied {\sc mepsa} to an initial sample of 2024 light curves
in the full passband. We applied the same selection on the peak significance
and selected the long GRBs by requiring $T_{90}>2$~s, where $T_{90}$ was taken
from the GRB
catalog\footnote{http://gammaray.msfc.nasa.gov/batse/grb/catalog/current/tables/duration\_table.txt}
\citep{Paciesas99}. We ended up with a sample of 1089
long GRBs with at least two $5$--$\sigma$ significant peaks.
Overall we collected 7649 peaks and 6560 WTs.
Hereafter, this will be referred to as the BATSE sample.

We also applied the same selection procedure to the light curves corresponding to
the sum of the two softest energy channels (1 and 2) and to the sum of the two hardest channels
(3 and 4), respectively within the 25--110~keV and $>$110~keV bands. We collected 1065 and 922 GRBs,
with 5156 and 4912 WTs, respectively.
These two groups will be hereafter referred to as BATSE12 and BATSE34 sets, respectively.

\subsubsection{{\em Fermi}/GBM data}
\label{sec:GBMsample} 
We selected 586 long GRBs detected with {\em Fermi}/GBM \citep{Meegan09} from July 2008 to December 2013.
We extracted the light curves of the two brightest GBM units in the energy band 8--1000~keV with
64 ms resolution and subtracted the background through interpolation with a polynomial of up to
third degree. We selected the long GRBs by imposing $T_{90}>2$~s, where $T_{90}$ was taken from
the official catalog.\footnote{http://heasarc.gsfc.nasa.gov/W3Browse/fermi/fermigbrst.html}
We restricted to time intervals whose median values range from $-30$ to $300$~s with reference
to the trigger time. This corresponds to the time interval covered by the time tagged event (TTE)
data type in trigger mode \citep{Paciesas12,Gruber14}. Before $-30$~s and after 600~s the time resolution
is that of CTIME data, $0.256$~s. In most cases we did not consider intervals $t>300$~s, because
interpolation-estimated background often becomes critical and the required effort for a proper
estimate is beyond our scope \citep{Gruber11a,Fitzpatrick12}. We did not consider GRBs showing
prolonged activity beyond this time interval.
We then applied the same selection criteria as for the previous sets. Finally, by visual inspection
we removed phosphorescence spikes due to high--energy particles \citep{Meegan09}, by comparing
the same profiles in different GBM units. We ended up with a final sample of 2383 peaks out of
544 GRBs with at least two significant peaks. The total number of WTs is 1839.

\subsubsection{{\em Swift}/XRT X--ray flares}
\label{sec:XRTsample} 
We considered the catalog of 498 X--ray flare candidates detected with
{\em Swift}/XRT obtained by \citet{Swenson14}. This was extracted from
680 XRT light curves from January 2005 to December 2012 with a method
based on the identification of breakpoints in the residuals of the fitted
piecewise power--law light curves: these points mark sudden changes in the mean value
due to unfitted features. The optimal set of breakpoints was then found by
minimizing the residual sum of squares against piecewise constant functions.
To counter the effect of overfitting with unnecessary breakpoints, they made use
of the Bayesian Information Criterion (see \citealt{Swenson13,Swenson14} for further
details).
In this catalog each candidate is assigned a confidence value. We conservatively selected the
subsample with a minimum confidence of 90\%, ending up with 205 X--ray flare candidates.

We separately merged each X--ray flare catalog with the BAT one by joining, for
each GRB, the sequence of $\gamma$--ray peak times and flare peak times into
a unique sequence of temporal peaks. In doing this, every peak which was
seen in both instruments was tagged as a $\gamma$--ray peak and not considered
any more as an X--ray flare. Analogously to the requirements for the previous sets,
we selected those GRBs with at least two (either $\gamma$ or X--ray) peaks, so
as to have at least one WT.
We ended up with a sample of 1098 (954 $\gamma$-- and 144 X--ray)
peaks in 244 GRBs (01/2005 -- 12/2012). We hereafter refer to this joint set
as BAT-X sample.

Finally, we selected the subsample with known redshift, so as to derive
the WTD in the source rest--frame. This was done by simply correcting for cosmic dilation
and thus dividing the observed WTs by the corresponding $(1+z)$.
Unlike the width of a given pulse, which is affected
not only by cosmic dilation but also by the energy passband shift, for their nature
WTs are affected by the latter to a much lesser extent.
We found 359 WTs in 94 GRBs with known redshift.
The subset with known redshift will be referred to as BAT-Xz.

As an independent check, we in parallel considered the X--ray flare catalogs of \citet{Chincarini10}
and \citet{Bernardini11}, which respectively include 113 early--time ($t<10^3$~s) X--ray flares
from April 2005 to March 2008 and 36 late--time ($t>10^3$~s) flares from April 2005
to December 2009. However, due to lower statistics, we hereafter focus on the BAT-X sample.

\subsection{Waiting time distribution modeling}
\label{sec:model} 
In physics a Poisson process is usually assumed to be characterized by a constant expected rate.
The WTD of this process is exponential with e--folding $\tau=1/\lambda$,
\begin{equation}
P(\Delta t)\ =\ \frac{1}{\tau}\,{\rm e}^{-\Delta t/\tau}\ =\ \lambda\,{\rm e}^{-\lambda\, \Delta t}\;,
\label{eq:WTDexp}
\end{equation}
where $\lambda$ is the constant mean rate and $\tau$ is the mean WT.
A time--varying Poisson process is characterized by a variable mean rate $\lambda(t)$:
the process is locally Poisson, but the expected rate changes with time as described by $\lambda(t)$.
According to this definition, the resulting process is the combination of two different processes
at play and is often referred to as a ``Cox process'' (e.g., \citealt{CoxIsham80}):
\begin{description}
\item[(a)] at a given time $t$ events are generated according to a Poisson process with rate $\lambda=\lambda(t)$
  and, as such, are statistically independent;
\item[(b)] the expected rate $\lambda$ is itself a function of time, which can vary either randomly
  or deterministically as time passes.
\end{description}

To derive the corresponding WTD, one may approximate $\lambda(t)$ as a piecewise constant function
in a number of adjacent time intervals $t_i$ $(i=1,\ldots,n)$ and treat it as a sequence of
several Poisson processes with rate $\lambda_i$.
Following \citet{Aschwanden10} and references therein, the resulting WTD is
\begin{equation}
P(\Delta t)\ \approx\ \sum_i \phi(\lambda_i)\,\lambda_i\, {\rm e}^{-\lambda_i\, \Delta t}\;,
\label{eq:aschw1}
\end{equation}
where
\begin{equation}
\phi(\lambda_i)\ =\ \frac{\lambda_i\,t_i}{\sum_j \lambda_j t_j}\;,
\label{eq:aschw2}
\end{equation}
is proportional to the expected number of WTs in interval $t_i$
where $\lambda=\lambda_i$. In the continuous limit, Eq.~(\ref{eq:aschw1}) becomes
\begin{equation}
P(\Delta t)  =  \frac{\int_0^{T} \lambda(t)^2\,{\rm e}^{-\lambda(t)\,\Delta t}dt}{\int_0^{T} \lambda(t)\,\,dt}\;,
\label{eq:aschw3}
\end{equation}
where $T$ is the total duration.
When $\lambda(t)$ is either unknown or hard to treat, it is possible to define $f(\lambda)$
such that $f(\lambda)\,d\lambda=dt/T$, that is the fraction of time during which the expected rate
lies within the range $[\lambda, \lambda+d\lambda]$. Equation~(\ref{eq:aschw3}) becomes,
\begin{equation}
P(\Delta t)  =  \frac{\int_0^{+\infty} f(\lambda)\,\lambda^2\,{\rm e}^{-\lambda\,\Delta t}d\lambda}{\int_0^{+\infty} \lambda\,f(\lambda)\,d\lambda}\;.
\label{eq:aschw4}
\end{equation}

We adopted the model for $f(\lambda)$ provided by \citet{Li14} in their eq.~(5),
which has been proposed to fit the WTD obtained for solar X--ray flares and solar energetic
particle events,
\begin{equation}
f(\lambda)\ =\ A\,\lambda^{-\alpha}\,\exp{(-\beta\,\lambda)}\;,
\label{eq:li14_lambda}
\end{equation}
with $\alpha$ and $\beta$ free parameters and $A$ is a normalization term ($0\le\alpha<2$).
This model generalizes several other models which had been put forward in the
same context \citep{Wheatland00,Aschwanden10}.
The mean rate $\bar{\lambda}$ is
\begin{equation}
\bar{\lambda}\ =\ \int_0^{+\infty} \lambda\,f(\lambda)\,d\lambda = A\,\beta^{\alpha-2}\,\Gamma(2-\alpha)\;,
\label{eq:li14_avlambda}
\end{equation}
where $\Gamma()$ is the gamma function.
From Equations~(\ref{eq:aschw4}-\ref{eq:li14_lambda}) the corresponding WTD is
\begin{equation}
P(\Delta t) =  (2-\alpha)\,\beta^{2-\alpha}\,(\beta+\Delta t)^{-(3-\alpha)}\;,
\label{eq:li14_WTD}
\end{equation}
and it is normalized like a probability density function (pdf), i.e.
$\int_0^{+\infty} P(\Delta t)\,d(\Delta t)=1$.
There are only two free parameters, $\alpha$, which determines the level of clusterization,
and the characteristic WT $\beta$ at which the WTD breaks: at
$\Delta t\gg \beta$, Eq.~(\ref{eq:li14_WTD}) becomes a power--law with an index of $(3-\alpha)$.

\begin{figure}[t]
\centering
\includegraphics[width=8.2cm]{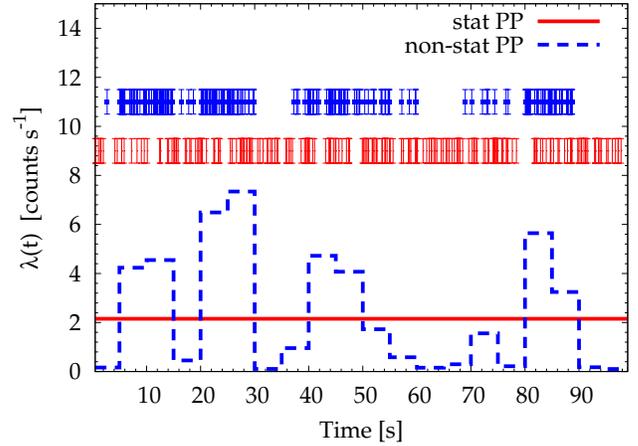}
\caption{Example of time--varying Poisson process with variable rate $\lambda(t)$
(dashed) assuming $\alpha=1$ and $\beta=0$ in Eq.~(\ref{eq:li14_lambda}) and a
constant one (solid) with the same mean values. Squares and vertical bars in the
top region mark the corresponding event times that were generated as a consequence.
The pronounced clusterization of the variable case over the constant one is clear.}
\label{fig:stat_vs_nonstat}
\end{figure}

Equation~(\ref{eq:li14_lambda}) naturally gives rise to clusterization, i.e. time intervals
characterized by an intense activity with a high rate of peaks (high $\lambda$), interspersed
with quiescent periods, during which the rate drops significantly (low $\lambda$).
The larger $\alpha$, the shallower the power--law regime at long WTs, and the more
clustered the time profile \citep{Aschwanden10,Aschwanden14}.
The details of how clustered the time profile looks like, in particular how the shot rate
varies with time, are directly described by Eq.~(\ref{eq:li14_lambda}).
At a given average rate $\bar{\lambda}$, by increasing $\alpha$ the variance of $\lambda$
increases correspondingly, i.e. the shot rate varies more wildly. This means deviating
more and more from the constant--rate case, thus enhancing the clustering character.
Figure~\ref{fig:stat_vs_nonstat} illustrates the difference between a time--varying
process like ($\alpha=1$, $\beta=0$) in Eq.~(\ref{eq:li14_lambda}) and a constant one 
sharing the same mean rate over a 100--s time window. The temporal sequence of events
for the former is evidently more clustered than that of the latter and, in spite of the
typical fluctuations of a Poisson point process, tracks the behavior of $\lambda(t)$.
It is worth nothing that, in general, in a Poisson process individual events are
independent of each other and, as such, have no memory of the events that occurred earlier,
{\em regardless} of whether the expected rate $\lambda$ is constant or time--dependent.
The difference instead lies in the observed average rates as a function of time, so not on (a)
but on (b):
depending on whether $\lambda(t)$ varies either in a deterministic way, or randomly with/without
memory, the average rate inherits the corresponding degree of correlation.

The WTDs we wanted to model are characterized by rare long WTs, where the count
statistics is so low that one cannot use a simple $\chi^2$ minimization to fit the expected
distribution of Eq.~(\ref{eq:li14_WTD}) to the counts collected in each bin.
On the other side, merging the bins so as to have enough counts loses information and
resolution. We therefore devised a log--likelihood based on Poisson statistics, which is
essentially the C statistic \citep{Cash79} and holds exactly even in the low count regime.
We used it in the context of a Bayesian Markov Chain Monte Carlo approach.
The details are reported in Appendix~\ref{sec:app}.

\section{Results}
\label{sec:res}
Table~\ref{tab:fit} reports the best fit parameters for all of the WTDs we considered.
In all cases the model of Eq.~(\ref{eq:li14_WTD}) provides an acceptable description.
The lowest confidence level is that of BATSE ($3.0$\%), still comparable with nominal 5\%
usually adopted as a threshold.
The BATSE sample is the largest (6560 WTs), so the high statistical
sensitivity is likely to enhance small deviations from the model.

\begin{figure}[t]
\centering
\includegraphics[height=9cm,angle=270]{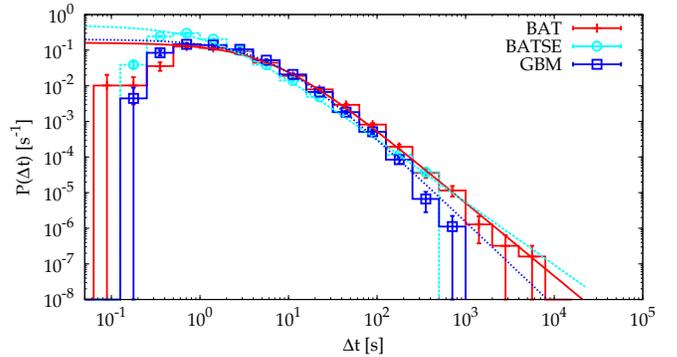}
\caption{WTDs of $\gamma$--ray pulses of the BAT (crosses), BATSE (circles), and GBM (squares)
samples together with their corresponding best fit models. Error bars are the (normalized) square
root of counts and are just indicative.}
\label{fig:nur_gamma}
\end{figure}
Figure~\ref{fig:nur_gamma} displays the WTDs for the $\gamma$--ray peak samples only.
Apart from the GBM, whose power--law index is significantly steeper, the BAT and BATSE
samples are fit with comparable indices, $2.06_{-0.09}^{+0.10}$ and $1.76\pm0.04$, respectively.
This is remarkable, given the different kind of detectors, energy passbands, different
GRB populations each instrument is mostly sensitive to \citep{Band06}.
The soft and hard BATSE samples have the same index, showing no significant dependence
on the energy channel.
\begin{table}
\begin{center}
\caption{Best fit parameters of the model in Eq.~(\ref{eq:li14_WTD}) obtained
for different WTDs. Size is the number of WTs.\label{tab:fit}}
\begin{tabular}{lrcccc}
\tableline\tableline
Sample & Size & $\alpha$ & $\beta$ & PL index & CL\\
       &      &          & (s)     & $(=3-\alpha)$ & (\%) \\
\tableline
BAT      & 1582 & $0.94_{-0.10}^{+0.09}$ & $6.53_{-0.98}^{+1.22}$ & $2.06_{-0.09}^{+0.10}$ & $26.4$\\
BATSE    & 6560 & $1.24\pm0.04$ & $1.53_{-0.16}^{+0.19}$ & $1.76\pm0.04$ & $3.0$\\
BATSE12  & 5156 & $1.19\pm0.05$ & $2.72_{-0.29}^{+0.33}$ & $1.81\pm0.05$ & $7.5$\\
BATSE34  & 4912 & $1.18\pm0.05$ & $1.23_{-0.16}^{+0.18}$ & $1.82\pm0.05$ & $76.6$\\
GBM      & 1839 & $0.64_{-0.17}^{+0.16}$ & $6.76_{-1.14}^{+1.44}$ & $2.36_{-0.16}^{+0.17}$ & $36.3$\\
BATtrunc & 1445 & $0.78_{-0.16}^{+0.15}$ & $6.99_{-1.28}^{+1.63}$ & $2.22_{-0.15}^{+0.16}$ & $5.2$\\
BAT-X    &  854 & $1.34_{-0.07}^{+0.06}$ & $6.33_{-1.20}^{+1.54}$ & $1.66_{-0.06}^{+0.07}$ & $5.4$\\
BAT-Xz   &  359 & $1.45_{-0.11}^{+0.10}$ & $1.26_{-0.42}^{+0.72}$ & $1.55_{-0.10}^{+0.11}$ & $18.5$\\
\tableline
\end{tabular}
\end{center}
\end{table}
We investigated the reasons for the steeper WTD of the GBM set as follows: the dearth of
long WTs is likely due to the shorter scanned time intervals, mostly from $-30$ to $300$~s
(Sect.~\ref{sec:GBMsample}). We therefore truncated the light curves of the {\em Swift}/BAT set
and revised the WT selection accordingly. The results are reported in Table~\ref{tab:fit} as
BATtrunc set, which includes 1445 WTs. Compared with the original BAT set, the WTD of the
truncated data becomes steeper, from $2.06_{-0.09}^{+0.10}$ to $2.22_{-0.15}^{+0.16}$, i.e.
compatible with the GBM value within uncertainties. Hence we interpret the slightly steeper
value of the GBM set as the result of shorter time profiles which disfavor long WTs.

\begin{figure*}
\centering
\includegraphics[height=17cm,angle=270]{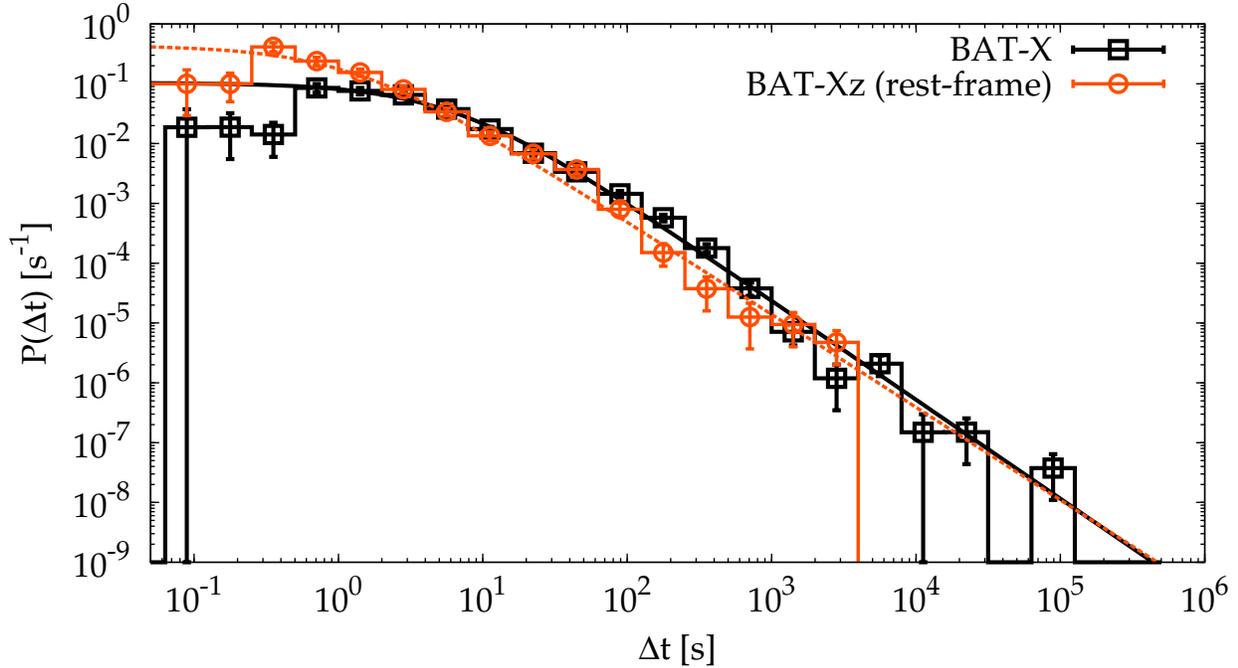}
\caption{WTDs of $\gamma$--ray pulses and X--ray flares of BAT-X (squares) and of BAT-Xz (circles)
sets together with their corresponding best fit models. Error bars are the (normalized) square root
of counts and are just indicative.}
\label{fig:nur_X_SR_mitZ}
\end{figure*}
Figure~\ref{fig:nur_X_SR_mitZ} displays the BAT-X set (squares) together with the corresponding best
fit model.
The power--law index is $1.66_{-0.06}^{+0.07}$, i.e. compatible with BATSE sets within uncertainties.
What is more, {\em merging X--ray flares did not change the stochastic nature exhibited by the
WTD, but extended its dynamical range by at least one order of magnitude with WTs as long
as $10^5$~s. A common stochastic model is found to well describe the WTD observed across
more than five orders of magnitude.}

A similar result is obtained when one restricts to the known--redshift sample BAT-Xz
in the GRB rest--frame (circles in Fig.~\ref{fig:nur_X_SR_mitZ}):
here the power--law index is $1.55_{-0.10}^{+0.11}$, i.e. somewhat
shallower. The rest--frame characteristic time is significantly shorter because of cosmic
dilation, $1.3$~s instead of the observer--frame values of 6--7~s.

We also searched for possible correlations between WTs and peak intensities and between WTs and
peak fluences of adjacent pulses, but found none.
Finally, we repeated the analysis for various subsets of GRBs, by requiring a minimum number of
of peaks per GRB and found no significant difference.

\subsection{$\gamma$--ray peaks vs. X--ray flares}
\label{sec:pulses_flares} 
\begin{figure}
\centering
\includegraphics[width=9cm]{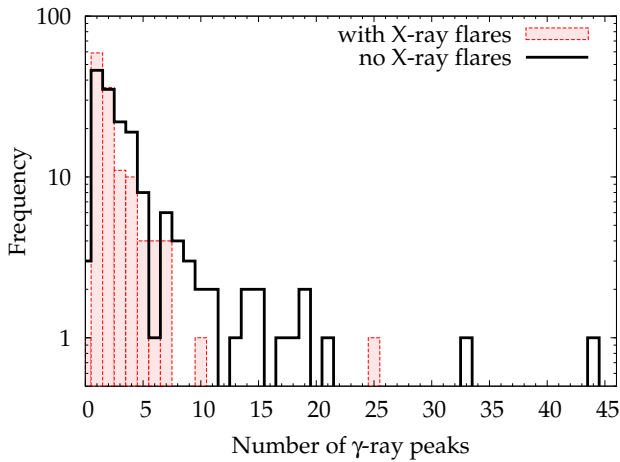}
\caption{Distribution of number of $\gamma$--ray peaks per GRB for two distinct subsets
of the {\em Swift} BAT-X sample, depending on the presence/lack of flares in the following X--ray
emission. Almost all (23/25) GRBs with at least 8 $\gamma$--ray peaks have no X--ray flares.
A KS test yields a common population probability of $0.21$\%.}
\label{fig:n_pulses_flares}
\end{figure}

Figure~\ref{fig:n_pulses_flares} shows the distribution of the number of $\gamma$--ray peaks
per GRB for two different classes of GRBs, depending on whether their subsequent X--ray emission
contains X--ray flares.
Surprisingly, it is found that almost all GRBs (23/25) with $N\ge8$ $\gamma$--ray peaks have no X--ray
flares, although the two groups have comparable size, 131 and 163 GRBs with and without flares,
respectively. The two distributions are unlikely to share a common population of events: a
Kolmogorov--Smirnov (KS) test yields a mere $0.21$\% probability, i.e. they are different with
$3.1$$\sigma$ (Gaussian) confidence.
We visually inspected each of these $\gamma$-- and X--ray light curves and found only one case of
a flareless GRB, whose X--ray light curve exhibited some low--level flaring activity which did not
pass the 90\%--confidence threshold in the flare sample selection (Sect.~\ref{sec:XRTsample}).
Therefore, GRBs with many pulses are unlikely to exhibit flares in the following declining X--ray emission.

\section{Discussion}
\label{sec:disc}
Our results may be summarized in four fundamental aspects:
\begin{enumerate}
\item $\gamma$--ray peaks and X--ray flares are compatible with being
different aspects of the same stochastic process, which goes on after
the end of the GRB itself and spans more than five orders of magnitude in time;
\item short (interpulse) and long (quiescent) WTs between $\gamma$--ray
peaks are different realizations of the same stochastic process, the latters being only less
frequent than the formers; hence a GRB with QTs and another without
are by no means more different from each other than any other kind of GRBs are;
\item GRBs with several ($\ge8$) $\gamma$--ray pulses are unlikely to exhibit X--ray
flares after the end of the prompt emission;
\item $\gamma$--ray peaks and X--ray flares tend to cluster in much the same
way that solar flares, energetic particle events, and CMEs do, even though the processes
may be different.
\end{enumerate}
The lognormal nature of the WTD originally claimed \citep{Li96} has recently been shown
to be possibly an artifact of the peak detection algorithms in the short WT end ($\la1$~s),
where peaks significantly overlap and can hardly be separated \citep{Baldeschi15}.
We found that the long value ($>$ few s) tail needs no more to be described as the sum
of a lognormal tail and a power--law excess due to the presence of QTs, that were
interpreted as a different component \citep{NakarPiran02a,Quilligan02,DragoPagliara07}.
This apparent diversity also concerns the so--called precursors
\citep{Lazzati05,Burlon08,Burlon09,Charisi14}, which are nothing but emission periods
that are less intense than the following activity from which they are separated by a quiescent
time.
Our results (1. and 2.) show that all kinds of WTs, including precursors,
can be described within a common stochastic process, and this holds all the way up
to late X--ray flares, thus pointing towards a common mechanism,
which keeps on working during and after the end of the prompt $\gamma$--ray emission, before
the afterglow emission due to the interaction with the external medium takes over.

Another question concerns the break at low values in the WTD modeled in terms of the
characteristic WT $\beta$: is it an intrinsic property or is it entirely due to the low
efficiency at short values of the peak detection algorithm?
The capability of separating overlapping structures depends on a number of variables,
such as the ratio between WT and peak widths, on intensities, and on temporal structures.
While the drop at $\Delta t\la0.5$~s is certainly due to the algorithm efficiency, the break
itself modeled with $\beta$ is more complex: $\beta$ is shorter at harder energies
(Table~\ref{tab:fit}).
A given pulse has a narrower temporal structure at harder energies \citep{Fenimore95}, whereas
in the softest energy channels there is a slow--varying component \citep{Vetere06}.
The presence of such soft component might hinder the peak identification in some case, so we
examined the light curves in the harder channels. Visual inspection suggests that the paucity
of subsecond WTs with respect to the power--law extrapolation is real and is unlikely to be a mere
artifact of the peak identification process.
In addition, minimum pulse widths observed in GRB profiles typically are in the range
$0.1$--$1$~s \citep{Fenimore95,Norris96,Margutti11c}. The MEPSA efficiency is above 10\%
for such pulse widths, for WTs $>0.5$~s, and for measured signal to noise ratios $>5$
\citep{Guidorzi15}.
It is therefore unlikely that the algorithm efficiency is entirely responsible for
the observed exponential cutoff observed in the low end of the WTDs.

\subsection{A simple toy model}
\label{sec:toy} 
We devised a very simple toy model to explore more in detail how a time--dependent
Poisson process like the one of Eq.~(\ref{eq:li14_lambda}) could be obtained in a GRB
engine. For the sake of clarity, suppose each pulse marks the accretion of a single
fragment of an hyperaccreting disk.
Actually, the idea behind this model is more general and only deals with the sequence
of bursty emission episodes and their probability of occurring within a given time; however,
hereafter we refer to the model of an hyperaccreting disk being fragmented as the source
of the stochastic process which is responsible for the prompt $\gamma$--ray emission
\citep{Woosley93,MacFadyen99} as well as for the subsequent X--ray flare activity
\citep{King05,Perna06,Kumar08,Cannizzo09,Geng13}.
We briefly summarize the basic ingredients of the model, which are then thoroughly described
in the remaining part of this Section:
\begin{itemize}
\item a number of fragments are independently accreted with the same, constant, probability
per unit time;
\item the number of available fragments is obviously decreasing with time; this naturally
leads to a time--dependent Poisson process whose mean accretion rate decreases with time;
\item at the beginning, if the mean rate is too high ($\lambda>1/\beta$),
accretion becomes inefficient and is suppressed by a factor of $\exp{(-\lambda\beta)}$;
\item for some ($\sim30$\%) GRBs the reservoir of fragments is split into two separate groups
sharing the same individual accretion probability per unit time, but with the second group
becoming available only at later times (e.g., the late group could be identified with the
outer part of the accretion disk).
\end{itemize}
Let us assume that the disk or the inner part of it has been split into a number of fragments,
each of which has the same given probability of accreting per unit time, independently of
the others. The probability for a given fragment to survive up to a given time $t$ is
proportional to $\exp{(-t/\tau)}$, where $\tau$ is the mean accretion time for each fragment.
The total expected rate scales as the number of fragments that are still available,
$\lambda=|\dot{N}(t)|=(N_0/\tau)\,\exp{(-t/\tau)}=N(t)/\tau$.
The analogous $f(\lambda)$ to Eq.~(\ref{eq:li14_lambda}) is found as
\begin{equation}
f(\lambda)\,d\lambda\ \propto |dt| = \tau\ \frac{d\lambda}{\lambda}\;,
\label{eq:toy1}
\end{equation}
which corresponds to the $\alpha=1$ case in Eq.~(\ref{eq:li14_lambda}) at $\lambda\ll1/\beta$.
\begin{figure}
\centering
\includegraphics[width=8.5cm]{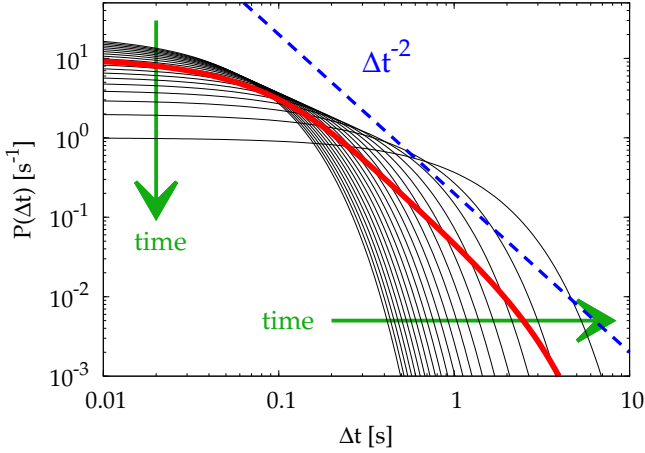}
\caption{Series of exponential WTDs (thin solid) of a sequence of constant Poisson processes
with progressively increasing e--folding. At intermediate values the mean WTD (thick solid)
scales as a power--law with index 2, shown for comparison (dashed).}
\label{fig:WTD_toy}
\end{figure}
Rather than a continuously varying, $\lambda(t)$ of this model is described more exactly by
a piecewise Poisson process like the one of Eqs.~(\ref{eq:aschw1}-\ref{eq:aschw2}),
where $\lambda_i=i/\tau$ is the expected rate when $i$ fragments are left over.
All terms have equal weights $\phi_i$'s, since each piecewise constant process contributes
one WT. The resulting WTD is thus given by Eq.~(\ref{eq:aschw1}),
\begin{equation}
P(\Delta t)\ = \frac{1}{N_0}\ \sum_{i=1}^{N_0} \frac{i}{\tau}\, {\rm e}^{-i\, \Delta t/\tau}\;,
\label{eq:toy2}
\end{equation}
which can also be expressed as,
\begin{equation}
P(x)\ = \frac{x\,\big[N_0\,x^{N_0+1} - (N_0+1)\,x^{N_0} + 1\big]}{N_0\,\tau\,(1-x)^2}\;,
\label{eq:toy3}
\end{equation}
where $x={\rm e}^{-\Delta t/\tau}$. In Figure~\ref{fig:WTD_toy} an example of such WTD is shown,
with $N_0=20$ initial fragments, $\tau=1$~s. As time goes by, $\lambda(t)$ decreases and the
e--folding of the individual exponential WTDs (thin solid) increase correspondingly. As a result,
the total WTD (thick solid) show a power--law regime with index 2 at intermediate values of $\Delta t$.
At $\Delta t\la \tau/N_0$ the total WTD is dominated by the initial exponential with e--folding
$\tau/N_0$, when fragments are all available. This agrees with the result that the intrinsic (i.e.,
corrected for the algorithm efficiency) WTD at short values is likely exponential, that is, compatible
with a constant Poisson process \citep{Baldeschi15}.

Thus far, with reference to Eq.~(\ref{eq:li14_lambda}) our model implicitly assumed $\beta=0$
(see Eq.~\ref{eq:toy1}). However,  in our attempt to reproduce the observed WTD with the piecewise
constant process of Eq.~(\ref{eq:toy2}) failed to model the observed break at $\Delta t\sim 1/\beta$.
So we required that, when the expected rate becomes comparable or higher than $1/\beta$, the number of observed
WTs is suppressed by a factor of $\exp{(-\lambda\beta)}$ with respect to our model.
This can be interpreted as if, when the number $N$ of fragments that can be readily accreted
is such that the expected rate is $\lambda=N/\tau\ga 1/\beta$, the overall process becomes inefficient
and the rate is suppressed by $\exp{(-\lambda\beta)}$. In other words, the number of WTs shorter
than $\beta$ is smaller than what is expected from Eq.~(\ref{eq:toy1}).
This introduces some degree of memory in the initial stages of the accretion process: as long
as the number of fragments ready to be accreted is too high ($\tau/N\la\beta$) some of them are
temporarily halted from accreting by some mechanism connected with the accretion rate itself.
For instance, this self--regulating mechanism could be driven by the magnetic field (e.g.,
\citealt{Proga06,Uzdensky06,Bernardini13}), which is known to have a complex role in accretion
processes of utterly different objects such as T~Tauri stars \citep{Stephens14}.
However, we cannot provide a more specific and physical justification for the exponential
character of this self--quenching mechanism, which is therefore ad--hoc in its present formulation.

We assumed the logarithmic average and 1-$\sigma$ scatter of the BAT-X WTD, $16.8$~s and a multiplicative
scatter of $7.1$,  to generate the values for $\tau$ for each simulated burst.
The number of generated peaks in each simulated curve was taken from the observed distribution and
was augmented by 20\% to ensure that the detected peaks were enough (since some are missed by
the algorithm).
The peak times for each simulated curve were randomly generated from an exponential distribution
with e--folding $\tau$, i.e. independently from each other. To mimic the drop in the peak detection
efficiency at short WTs as well as the mechanism mentioned above about the suppression at high rates,
we overlooked each peak occurring within $0.5$~s of the previous one, while through a binomial
we assigned each $\Delta t$ a probability $\exp{(-\beta/\Delta t)}$ of being observed, where
$\beta$ was set to the fitted value of the real WTD (Table~\ref{tab:fit}).

\begin{figure}
\centering
\includegraphics[height=9cm,angle=270]{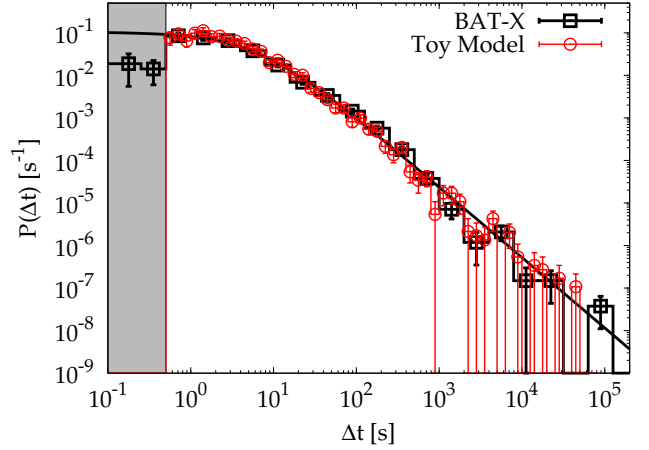}
\caption{WTD for the {\em Swift} BAT-X sample (squares) compared with a simulated sample of 903
WTs derived from a toy model (circles). The shaded interval is where the peak search algorithm
efficiency drops.}
\label{fig:DT_toy}
\end{figure}
To obtain a good match with the observed WTD over the same range we had to make a further assumption:
we assumed that for a fraction of GRBs ($\sim30$\%) randomly selected through a binomial, the disk is fragmented
equally in two groups, the first of which is available for being accreted from the beginning ($t_0=0$),
while the second one becomes available from $t_0=50\,\tau$ on, where $\tau$ is the common mean accretion
time for each individual fragment from $t=t_0$. The reason behind this is the observation of two similar
bunches of $\gamma$--ray peaks interspersed with a long quiescent time (up to several tens of seconds) for
a small fraction of GRBs. Physically, this could be the result of an outer part of the disk being accreted at
later times with respect to the inner one or, more in general, delayed additional energy reservoir
becoming available for late internal dissipation, with minimum variability timescale comparable with
that of the early prompt emission \citep{FanWei05,LazzatiPerna07,Troja14}.
While the choice of the fraction of such GRBs and the duration of the quiescence period
are somewhat arbitrary, the good match between simulated and real WTD does not depend crucially on them.
Overall, the goal here is just to show the plausibility of the essential properties of this model, which
can reproduce the observed properties in spite of the simple assumptions.
We ended up with a set of 903 simulated WTs, whose distribution is compared with the real one
in Fig.~\ref{fig:DT_toy}.

\begin{figure}
\centering
\includegraphics[width=8.7cm]{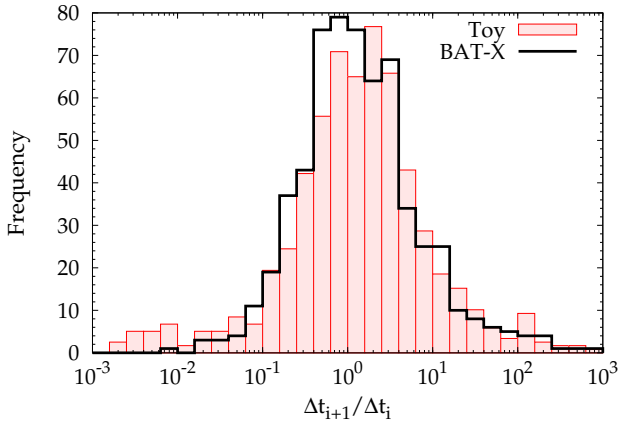}
\caption{Distribution of the ratio of adjacent WTs of the {\em Swift} BAT-X sample (solid)
and of the toy--model sample (shaded).}
\label{fig:logratio}
\end{figure}
We further tested the consequences of this toy model by studying the distribution of the ratio
between adjacent WTs. While WTDs describe how WTs distribute as a whole, losing information on their
temporal sequence, the ratio distribution focuses on that. We therefore selected from the BAT-X
as well as from the simulated sample the GRBs with $\ge3$ peaks, so as to have at least two WTs,
and derived the two distributions shown in Fig.~\ref{fig:logratio}.
A KS test between the two sets yields 43\% probability that they were drawn from a common population.
Logarithmic mean and dispersion for the real (simulated) data are $\mu=0.14$ and $\sigma=0.72$ ($\mu=0.09$
and $\sigma=0.81$). Similar results are obtained adopting other non--parametric tests, such as the
Wilcoxon--Mann--Whitney or the more sensitive Epps--Singleton one \citep{EppsSingleton86},
respectively yielding 45\% and 9\% probability.
Interestingly, simply replacing Eq.~(\ref{eq:toy1}) with a constant Poisson process and applying the
very same following steps, one ends up with a narrower and more zero-centered logarithmic ratio
distribution, $\mu=0.013$ and $\sigma=0.32$, which is rejected with high confidence (p--value $<2\times10^{-16}$)
from a KS test. This means that for a constant process the ratio is,
on average, closer to 1, and is less scattered around it than real data.
The compatibility of the ratio distribution predicted by the toy model with the real one proves that
the temporal sequence of WTs is compatible with an evolving Poisson process and is incompatible with
a constant one on long timescales.
In particular, X--ray flares are nothing but some of the last fragments that are left over and which
are accreted on long timescales, when the rate decreases in a granular way, following the very same
stochastic process ruling the accretion of the earliest ones. Hence, no correlation is to be
expected between $\gamma$--ray prompt emission duration ($T_{90}$) and X--ray flares times, 
in agreement with observations \citep{Liang06}.
Finally, the result of Sect.~\ref{sec:pulses_flares} can be easily explained: GRBs with many
$\gamma$--ray peaks accrete fragments rapidly with relatively short $\tau$, so that at late time
very few or none at all are left over for X--ray flares.
The same result could be explained differently though: multi--peaked GRBs could have on average
a brighter early X--ray afterglow continuum which overshines possible X--ray flares, which would then
go unseen.

Overall, we assumed a direct connection between emission and observed times of the peaks.
Within the context of internal shocks the observed time profile of both prompt $\gamma$--rays
and late X--ray flares is strictly connected to the emission history \citep{Kobayashi97,Maxham09}.
Should this not be the case, little could be inferred about on emission times -and potentially
the times of individual accretion episodes- from the study of the observed WTD.
However, this connection becomes more complex due to the variety of Lorentz factors associated
with the wind of shells colliding with each other. Even though the intrinsic duration of
the GRB engine activity may differ by a factor of a few from the observed one \citep{Gao14},
on average the temporal sequence of mutual collisions between randomly--assigned
Lorentz--factor shells should track the emission time history. The nature of
a given WTD is not altered as a whole when one passes from the emission to the observed times:
in fact, each shell has a Lorentz factor -which is in principle what can make the observed WTs
very different from the emitted ones- that is independent of the WTs preceding and following that shell.
This statistical independence ensures that the observed WTD keeps memory on the emission
time distribution. Only at late times, when the average Lorentz factor is expected to
systematically decrease and the statistical independent character likely begins to fail,
long WTs are likely to be affected as a consequence.

\subsection{Solar activity: analogies and differences}
\label{sec:sun} 
It is remarkable and intriguing that WTDs of both solar eruptive events (X--ray flares, radio storms,
high--energy particle events, CMEs) and of GRBs can be modeled with the same kind of
time-dependent Poisson process.
The power--law characterization of the WTD heavy tail must not be overinterpreted from a mathematical
viewpoint, since power--laws are, in general, what one ends up with when dealing with the sum of
independent heavy--tailed variables. It works much in the same way that a normal distribution is the
final outcome of the sum of independent finite--variance variables. In addition, claiming that data
are power--law distributed is contrived whenever the explored range does not cover at least two
decades \citep{Stumpf12}.
In this sense, invoking a SOC--driven mechanism for GRBs purely based on the power--law
character of the WTD, and possibly of the energy distribution too, as suggested for X--ray flares
from GRBs \citep{WangDai13} or from other black hole systems \citep{Wang14}, is a stretched
interpretation of the data, as we explain below.

The same or very similar power--law indices imply that both processes have
very similar degrees of clusterization, with analogous swings between intense and low--activity
periods, apart from temporal rescaling (seconds for GRBs, hours for the Sun).
However, one has to be careful extending this analogy to a common physical mechanism.
Overall, there is a fundamental difference in terms of dynamical systems between GRB inner engines
during core collapse and the Sun: for the latter, the regions where eruptive phenomena
take place continuously receive energy, which is then released through avalanches, thus making
the SOC interpretation plausible (although alternatives based on MHD turbulence seem equally
compatible with observations). Instead, GRB engines are systems which start with a
very-far-from-equilibrium configuration, evolve very fast using up all of the available energy, which
-no matter how much- is {\em limited}. A GRB inner engine cannot return
to its original configuration, it goes through an obviously {\em irreversible} evolution, whereas
this is not the case for the solar active regions over sufficiently long timescales.
For this reason, one needs not invoke SOC mechanisms related to accretion disks, in particular
there is no need for a mechanism like the one proposed to explain $1/f$ fluctuations in
black hole power spectra \citep{Mineshige94}.

Therefore, a simple time--varying Poisson process explains the secular evolution of
the mean rate of bursts/flares as well as the stochastic independent character of
individual energy release episodes. This model disregards the physical origin
of fragmentation and how energy is distributed among different fragments: thus, in principle
it is compatible with various physical drivers, such as
gravitational \citep{Perna06}, or magnetorotational instability \citep{Proga06}.


\section{Conclusions}
\label{sec:conc}
In this paper we studied the waiting time distributions of GRB $\gamma$--ray pulses
in three catalogs, {\em CGRO}/BATSE, {\em Fermi}/GBM, and {\em Swift}/BAT. For the
latter, for the first time we merged $\gamma$--ray pulses and X--ray flares detected
with {\em Swift}/XRT belonging to the same GRBs, and for a subsample the same analysis
was carried out in the source rest--frame.
We found that all WTDs can be described in terms of a common time--varying Poisson process
that rules different waiting time intervals, which thus far in the literature have
been treated differently: specifically, we showed that short WTs ($\la1$~s), long
quiescent times ($\ga10$~s) all the way up to late time X--ray flares are the manifestation
of a common stochastic process.
GRB WTDs exhibit heavy tails which are modeled with power--laws over 4--5 decades in time
with indices in the range $1.7$--$2.1$, depending on the relative weight of late time
events, such as X--ray flares, in each GRB sample. Because of the ubiquitous nature of
power--laws (central limit theorem for heavy--tailed distributions), the character of WTDs must
not be imbued with a mystical sense or overinterpreted as evidence for a universal process.
In this sense, the similarity of the WTD power--law index with that of solar eruptive phenomena,
such as flares and coronal mass ejections, proves nothing but a similar degree of clusterization in
time.
Nonetheless, it is remarkable that the WTD of $\gamma$--ray pulses and that of X--ray flares not only
have compatible power--law indices but they join and extend the dynamical range for a common sample of GRBs.
All this points to a common stochastic process ruling both phenomena.
The unification under a common process of all different kinds of waiting times in GRBs (short interpulse
times, long quiescent times, time intervals following precursors, time intervals between the end of the
$\gamma$--ray prompt emission and subsequent X--ray flares) is a new result.

Another noteworthy result is that GRBs with many ($\ge8$) $\gamma$--ray pulses are
unlikely ($3.1 \sigma$ confidence in Gaussian units) to exhibit X--ray flares in
their subsequent early X--ray emission. This result is naturally explained in
the context of the time--varying Poisson process: many pulses observed in the prompt
of a given GRB are indicative of a relatively short mean accretion time for a single
disk fragment. Consequently, most of the available fragments are consumed during
the prompt phase, with very few or none at all left over for the subsequent phase.

In the light of the irreversible evolution of GRB inner engines, the interpretation of
a time-varying Poisson process appears to be simple and reasonable: the secular
evolution of the expected rate of events is naturally linked to the energy reservoir
being gradually used up, whereas the stochastically independent accretion of individual
fragments is explained by the Poissonian character of the process.

Although self--organized criticality models naturally predict power--law tailed
distributions of waiting times and energy, drawing upon this kind of dynamics for GRBs
might be premature. Other equally plausible alternatives, such as fully developed
MHD turbulence, can explain the same properties, as it was suggested for the solar case.
Possible evidence for turbulence in GRBs has also been suggested from the analysis
of power density spectra \citep{Beloborodov98,Beloborodov00a,Guidorzi12,Dichiara13a}.
Yet, we find that a simple time--varying Poisson process such as that of a system
gradually using up all the available pieces already provides a remarkably accurate
description.

The energy distribution, which was beyond the scope of this paper, will help to further
constrain the stochastic process and possibly clarify whether more complex dynamical models,
such as SOC or MHD turbulence, are to be considered.

\acknowledgments 
We are grateful to the anonymous referee for a constructive and insightful review.
PRIN MIUR project on ``Gamma Ray Bursts: from progenitors to physics of the prompt
emission process'', P.~I. F. Frontera (Prot. 2009 ERC3HT) is acknowledged.

\appendix
\section{Log--likelihood to fit the distributions}
\label{sec:app}

Let the WTD consist of $N$ logarithmically spaced bins, each collecting $C_i$ counts.
Let $\Delta t_{i,1}$ and $\Delta t_{i,2}$ be the lower and upper bounds of the $i$--th bin
($i=1,\ldots,N$). Integrating Eq.~(\ref{eq:li14_WTD}) within this time interval yields
the corresponding expected counts, $E_i(\alpha,\beta)$, where we explicitly
meant that it depends on the model parameters:
\begin{equation}
E_i(\alpha, \beta)\ =\ C_{\rm tot}\ \int_{\Delta t_{i,1}}^{\Delta t_{i,2}} P(\Delta t)\ d(\Delta t)\ =\
C_{\rm tot}\ \beta^{2-\alpha}\ \Big[(\beta+\Delta t_{i,1})^{\alpha-2} - (\beta+\Delta t_{i,2})^{\alpha-2}\Big]
\label{eq:A1}
\end{equation}
where $C_{\rm tot}=\sum_{i=1}^N C_i$, and $\beta$ is a function of both model parameters (Sect.~\ref{sec:model}).
The probability of $C_i$ counts is ruled by the Poisson distribution where $E_i$ is the expected value,
\begin{equation}
P(C_i|\alpha, \beta)\ =\ \rm{e}^{-E_i}\ \frac{E_i^{C_i}}{C_i!}\;,
\label{eq:A2}
\end{equation}
where the dependence on model parameters is implicit through $E_i$. The total probability is thus
\begin{equation}
P({\bf C}|\alpha, \beta)\ =\ \prod_{i=1}^{N} \rm{e}^{-E_i}\ \frac{E_i^{C_i}}{C_i!}\;,
\label{eq:A3}
\end{equation}
where ${\bf C}$ is the set of observed counts per bin $\{C_i\}$ ($i=1,\ldots,N$).
The corresponding negative log--likelihood is therefore
\begin{equation}
L(\alpha, \beta)\ =\ -\sum_{i=1}^N \log{(P(C_i|\alpha,\beta))}\ =\
\sum_{i=1}^N \Big(E_i + \log{(C_i!)} -C_i\,\log{E_i} \Big)\;.
\label{eq:A4}
\end{equation}
We determine the best fit model parameters and their uncertainties in the Bayesian context.
The posterior probability density function of the parameters for a given observed distribution
${\bf C}$, is (Bayes theorem)
\begin{equation}
P(\alpha,\beta|{\bf C}) = \frac{P({\bf C}|\alpha,\beta)\ P(\alpha,\beta)}{P({\bf C})} ,
\label{eq:A5}
\end{equation}
where the first term in the numerator of the right-hand side of eq.~(\ref{eq:A5}) is
the likelihood function of Eq.~(\ref{eq:A3}), $P(\alpha,\beta)$ is the prior
distribution of the model parameters, and the denominator is the normalization term.
We assumed a uniform prior distribution, since no a priori information is available
on the model parameters. The mode of the posterior probability of Eq.~(\ref{eq:A5})
is therefore found by minimizing Eq.~(\ref{eq:A4}).

We estimate the posterior density of the model parameters through a Markov Chain Monte Carlo
(MCMC) algorithm such as the random--walk Metropolis--Hastings in the implementation of
the $R$\footnote{http://cran.r-project.org/} package {\sc MHadaptive}\footnote{
http://cran.r-project.org/web/packages/MHadaptive/index.html.} (v.1.1-8).
We initially approximate the posterior using a bivariate normal distribution centred on the
mode and with covariance matrix obtained by minimizing Eq.~(\ref{eq:A4}).
For each WTD we generate $5.1\times10^4$ sets of simulated model parameters and retain
one every 5 MCMC iterations after excluding the first 1000.
The remaining $10^4$ sets of parameters are therefore used to approximate the posterior
density. Finally, once the best fit model parameters are determined, the bivariate posterior
distribution of $(\alpha, \beta)$ is sampled via MCMC simulations, which yield
expected value and 90\% confidence intervals for each of them.

As a matter of fact, since the bins in the low end of distribution are strongly affected
by the poor efficiency of {\sc mepsa}, these are to be ignored. In practice, one has to
replace in Eq.~(\ref{eq:A1}) $C_{\rm tot}$ with $C'_{\rm tot}=\sum_{i=k_1}^{k_2} C_i$, where $k_1$
and $k_2$ are the first and last bins to be considered.
In addition, the same $E_i$ in Eq.~(\ref{eq:A1}) has to be further divided by a
renormalizing factor so that it becomes,
\begin{equation}
E_i(\alpha, \beta)\ =\ C'_{\rm tot}\ \frac{(\beta+\Delta t_{i,1})^{\alpha-2} - (\beta+\Delta t_{i,2})^{\alpha-2}}{(\beta+\Delta t_{k_1,1})^{\alpha-2} - (\beta+\Delta t_{k_2,2})^{\alpha-2}}\;.
\label{eq:A6}
\end{equation}
For the WTDs discussed in the present paper we considered $\Delta t\ge 0.5$~s
($\Delta t\ge 0.2$~s) in the observer (source) rest frame.

To assess the goodness of the fit for a given WTD, we use each set of simulated values
for $(\alpha, \beta)$ to generate as many synthetic WTDs from the the posterior predictive
distribution. Hence for a given observed WTD, we directly calculate $10^4$ synthetic realizations
of the same WTD. For each of these WTDs we calculate the negative log--likelihood with
Eq.~(\ref{eq:A4}) and derive a corresponding distribution of values, against which the value
obtained from the real WTD is checked. This comparison directly provides a confidence level
of modelling the observed WTD in terms of the best fit model of Eq.~(\ref{eq:li14_WTD}).

\end{document}